\documentclass{appolb}
\usepackage{graphicx}
\begin{document}
\title{Comparing conserved charge fluctuations from lattice QCD to HRG model calculations%
\headtitle{Comparing conserved charge fluctuations....}
\headauthor{Jishnu Goswami}
\thanks{Presented at Criticality in QCD and the Hadron Resonance Gas, 29-31 July 2020, Wroclaw Poland}%
}
\author{Jishnu Goswami \thanks{speaker}, Frithjof Karsch, Christian Schmidt
\address{Fakult\"at f\"ur Physik, Universit\"at Bielefeld, D-33615 Bielefeld,
	Germany}
\\
\vspace{4 mm}
{Swagato Mukherjee and Peter Petreczky
\address{Physics Department, Brookhaven National Laboratory, Upton, NY 11973, USA}
}
}
\maketitle
\begin{abstract}
We present results from lattice QCD calculations for $2^{nd}$ and $4^{th}$ order cumulants of conserved charge fluctuations and correlations, and compare these with various HRG model calculations. We show that differences 
between HRG and QCD calculations already show up in the second order cumulants close to the pseudo-critical 
temperature for the chiral transition in (2+1)-flavor QCD and quickly grow large at higher temperatures. 
We also show that QCD results for strangeness fluctuations are enhanced over HRG model calculations which are
based only on particles
listed in the Particle Data Group tables as 3-star resonances. This suggests the importance of contributions
from additional strange hadron resonances.
We furthermore argue that additional (repulsive) interactions, introduced either through excluded volume 
(mean field) HRG models or the S-matrix approach, do not improve the quantitative agreement with 
$2^{nd}$ and $4^{th}$ order cumulants calculated in lattice QCD. HRG based approaches fail to describe the 
thermodynamics of strongly interacting matter at or shortly above the pseudo-critical temperature of QCD.
\end{abstract}
  
\section{Introduction}
The theory of strong interactions, Quantum chromodynamics (QCD), also describe the thermodynamics of strongly interacting matter at finite temperature and density. It now is understood that at vanishing net baryon-number 
density the transition from low to high temperature reflects the physics of a true phase transition that 
occurs at vanishing values of the two light (up and down) quark masses in QCD and is due to the restoration
of chiral symmetry, which is spontaneously broken in the QCD vacuum and at low temperatures \cite{review}. 
QCD with its physical spectrum of light and strange quark masses undergoes a smooth transition from
hadronic bound states to the quark gluon plasma (QGP) at high temperature. 

Hadron Resonance Gas (HRG) models can be used to describe the thermodynamics of QCD at low temperature where the 
degrees of freedom of QCD matter are hadrons. This model assumes that interactions among hadrons can be accounted for by production of hadronic 
resonances which are added to thermodynamics as additional particles.  
The simplest implementation of a
non-interacting HRG model considers a mixture of ideal Bose gases for mesons and ideal Fermi gases for baryons. The total pressure of a hadronic medium then is obtained as the sum over individual contributions of partial pressures 
of different particle species. 
The HRG model can be justified using the S-matrix based virial expansion \cite{Dashen:1969ep}.
It has been shown that using partial wave analysis of the experimental scattering data non-resonant repulsive and attractive
interactions of hadrons largely cancel out in the thermodynamic quantities, and thus, the interactions can be indeed well
described by hadronic resonances \cite{Venugopalan:1992hy}. This approach has been recently revisited
in several papers \cite{Huovinen:2017ogf,Lo:2017lym,Fernandez-Ramirez:2018vzu}, where also some of the non-resonant (repulsive)
interactions were included.
%listed in Particle Data Group (PDG) booklet. 

HRG models have been used to extract information on thermal conditions at the time of freeze-out of hadrons 
from a high temperature partonic medium from experimental data on hadron yields \cite{Stachel:2013zma}.
However, a comparison of lattice QCD calculations of conserved charge
fluctuations with corresponding HRG model calculations shows that the latter deviates from QCD results more and 
more with increasing temperature and deviations are larger for higher order cumulants. This short-coming of
simple, non-interacting HRG models has been attempted to compensate by either taking into account further
contributions from repulsive interactions through thermodynamic calculations with extended hadrons 
\cite{Yen:1997rv} or with a repulsive mean field \cite{Huovinen:2017ogf}, since a comprehensive treatment
of the repulsive interactions in the S-matrix approach is not yet available.
While these modifications of point-like, non-interacting HRG model calculations generically lead to
a reduction of cumulants of conserved charge fluctuations, there also is evidence that strangeness fluctuations
calculated in QCD are larger than those obtained in HRG model calculations based only on experimentally
well established (3-star resonances) hadrons listed by the Particle Data Group (PDG). One popular approach to address
this issue is to include in HRG model calculations additional strange hadron resonances which are not listed in the 
PDG tables \cite{Bazavov:2014xya,Bazavov:2017dus,Pasztor:2018ipn}, but are obtained in quark model calculations \cite{Isgur,Ebert}. 
This too may be interpreted as an attempt to take care of further interactions.

We will present here a comparison of various cumulants of conserved charge fluctuations, calculated in 
$(2+1)$-flavor lattice QCD \cite{Bazavov:2014xya,Bazavov:2017dus}, with different HRG model calculations. We mainly focus on the temperature range close to the transition temperature which is relevant for setting the baseline for heavy-ion collision experiments. In the following we refer to the standard non-interacting HRG as PDG-HRG, where we consider down to 3 star hadrons and hadron resonances listed in PDG2020 \cite{Zyla:2020zbs}. 
We also extended the HRG model based on resonances listed by the PDG by using 
additional hadronic resonances obtained in relativistic quark model calculations 
(QM-HRG \cite{Isgur,Ebert}). 
Furthermore, we discuss modifications of the non-interacting HRG models obtained by including further interactions between  hadrons either through excluded volume  or mean field effects (EV-HRG) or an advanced treatment of the S-matrix 
approach to the thermodynamics of strongly interacting hadrons.

\section{Second order cumulants of conserved charge fluctuations and correlations}
Here we will discuss second order cumulants of net baryon-number ($B$) and strangeness ($S$) fluctuations. 
In particular we will compare the second order cumulants to various HRG model 
calculations and discuss to what extent deviations from HRG model results show up already on the level
of these low order cumulants.

\begin{figure}[b]
\includegraphics[width=4.3cm]{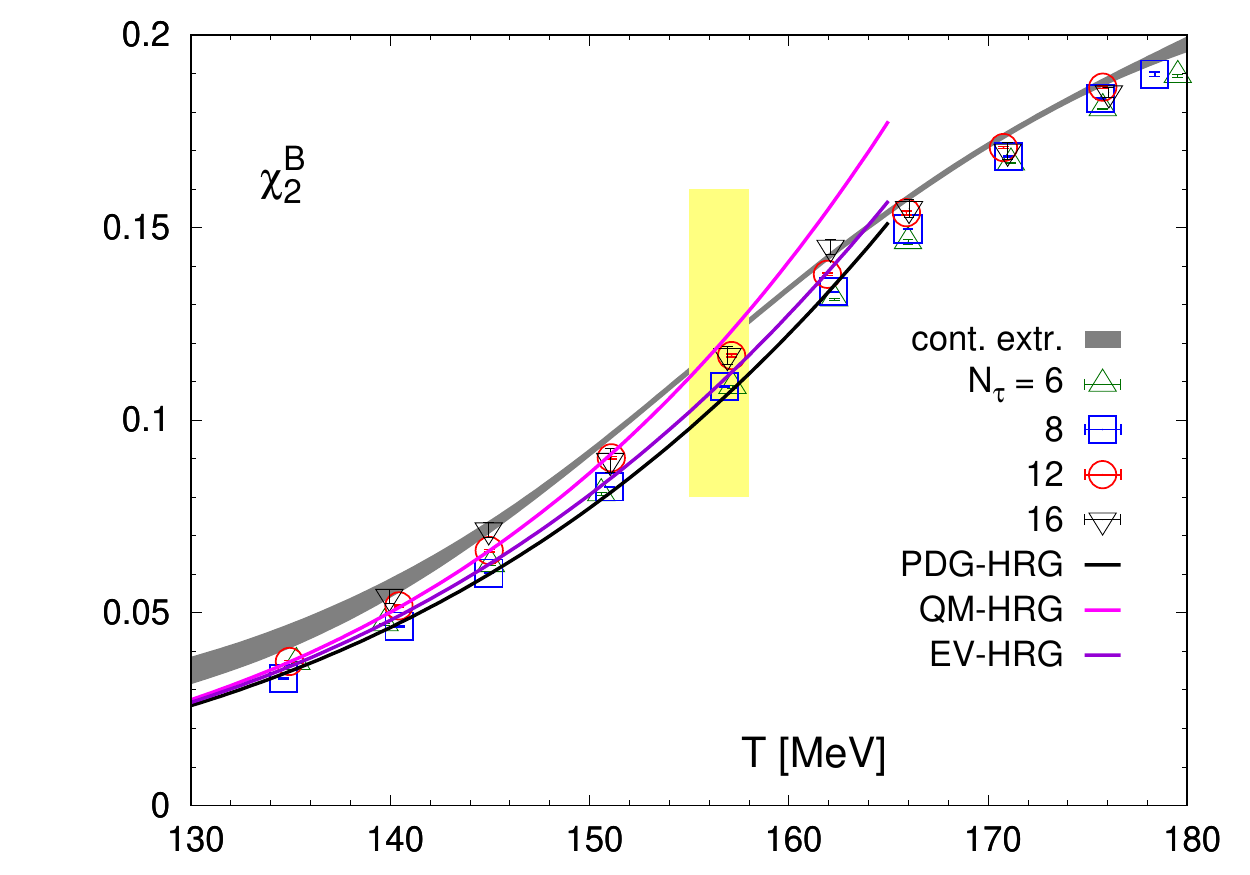}\hspace{-0.3cm}
\includegraphics[width=4.3cm]{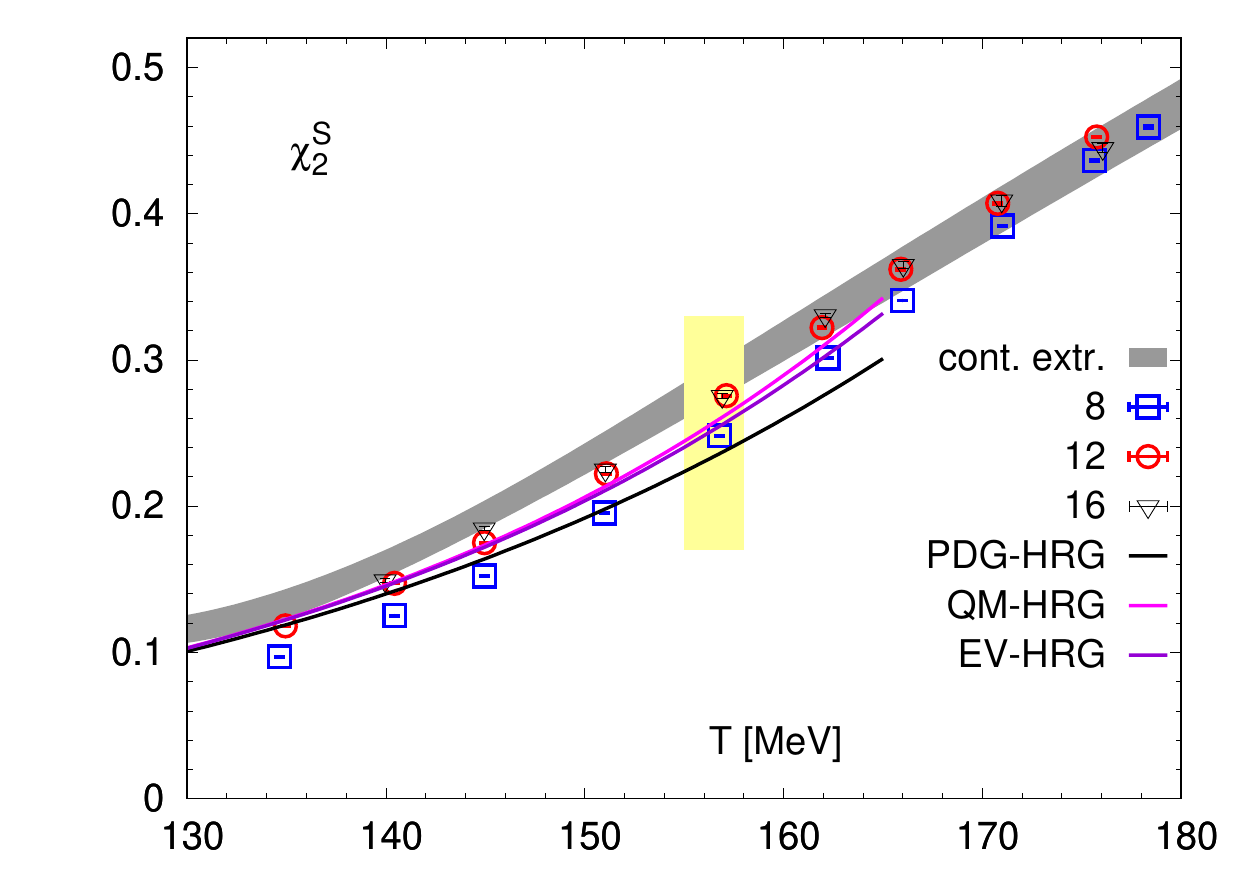}\hspace{-0.3cm}
\includegraphics[width=4.3cm]{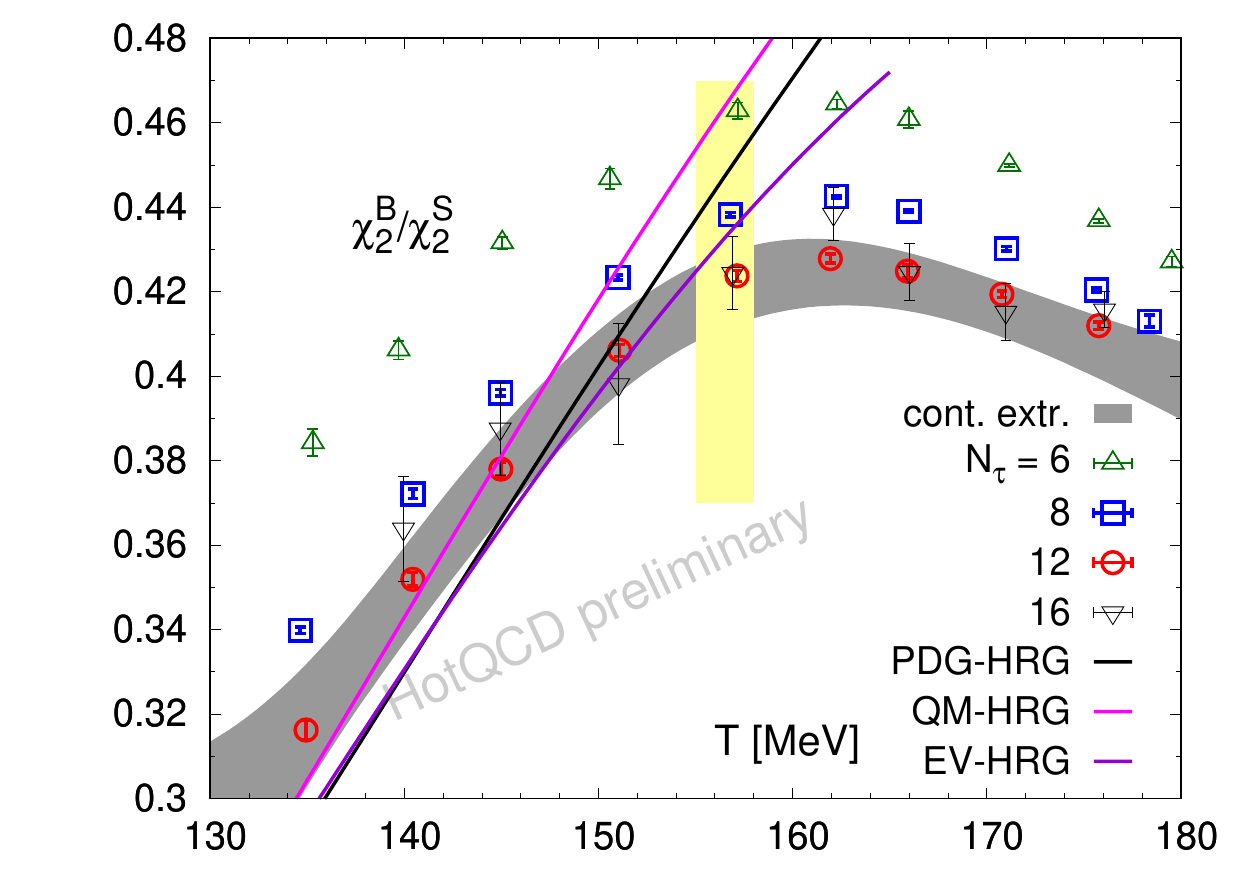}
\caption{Second order net baryon-number (left) and strangeness (middle) fluctuations as well as their
ratio (right).}
\label{fig:2nd_baryon_number}
\end{figure}

The pressure of the QCD partition function can be written as,
\begin{eqnarray}
P/T^4=\frac{1}{VT^3}\ln Z(V,T,\mu_B,\mu_Q,\mu_S).
\label{eq:qcd_pressure}
\end{eqnarray}
Generalized susceptibilities, {\it i.e.} the cumulants of conserved charge fluctuations, can be obtained by taking 
derivatives of the pressure with respect to baryon ($\mu_B$), 
electric charge ($\mu_Q$) and strangeness  ($\mu_S$) chemical potentials at  $\vec{\mu}=(\mu_B,\mu_Q,\mu_S)=0$,
\begin{eqnarray}
\chi_{lmn}^{BQS} &=&\frac{\partial^{l+m+n} P/T^4}{\partial (\mu_B/T)^l 
\partial (\mu_Q/T)^m \partial(\mu_S/T)^n}\bigg|_{\vec{\mu}=0} \nonumber \label{eq:gen_sus}. \\
\end{eqnarray} 

\subsection{Charge fluctuations}
In Fig.~\ref{fig:2nd_baryon_number}~(left, middle) we show results for net baryon-number ($\chi_2^B$) and
strangeness ($\chi_2^S$) fluctuations ($2^{nd}$ order cumulants). Results obtained on lattices of size
$N_\sigma^3\times N_\tau$, with $N_\sigma=4 N_\tau$ in (2+1)-flavor QCD simulations using the HISQ action
\cite{Bazavov:2017dus,Bazavov:2018mes}, 
are shown for several values of the lattice spacing, {\it i.e.} several values of temporal lattice extent 
$aN_\tau=1/T$. These data have been extrapolated to the continuum limit using a quadratic ansatz for
discretization errors in $(aT)$. As discussed above we compare these results to HRG model 
calculations based on hadron spectra listed in the PDG and obtained in quark model calculations,
respectively. Some basic formulas for the HRG model calculations are given in Appendices \ref{AA} and \ref{AB}. 

It is quite evident for $\chi_2^B$ and $\chi_2^S$ that PDG-HRG curves provide only a poor description of 
the QCD results close to the transition region. The agreement of HRG model and QCD results improves when one 
includes additional strange baryon resonances in the spectrum that are predicted in quark model calculations
(QM-HRG). However, as can be seen clearly in Fig.~\ref{fig:2nd_baryon_number}~(right), the non-interacting
HRG model calculations do not give a reasonable description of  the QCD results at temperatures
above the pseudo-critical temperature for the chiral transition, $T_{pc}=(156.5\pm 1.5)$~MeV \cite{Bazavov:2018mes}. 
In particular,
HRG results for the ratio  $\chi_2^B/\chi_2^S$ continue to rise above $T_{pc}$ while the QCD results
have a maximum close to $T_{pc}$ and then drop towards the non-interacting quark gas value, 
$(\chi_2^B/\chi_2^S)_{T\rightarrow \infty}=1/3$. Similarly, it is apparent that temperature derivatives 
of the $2^{nd}$ order cumulants keep rising in HRG model calculations while they reach a maximum for the 
QCD results close to $T_{pc}$.

\begin{figure}[t]
\centerline{
\includegraphics[width=5cm]{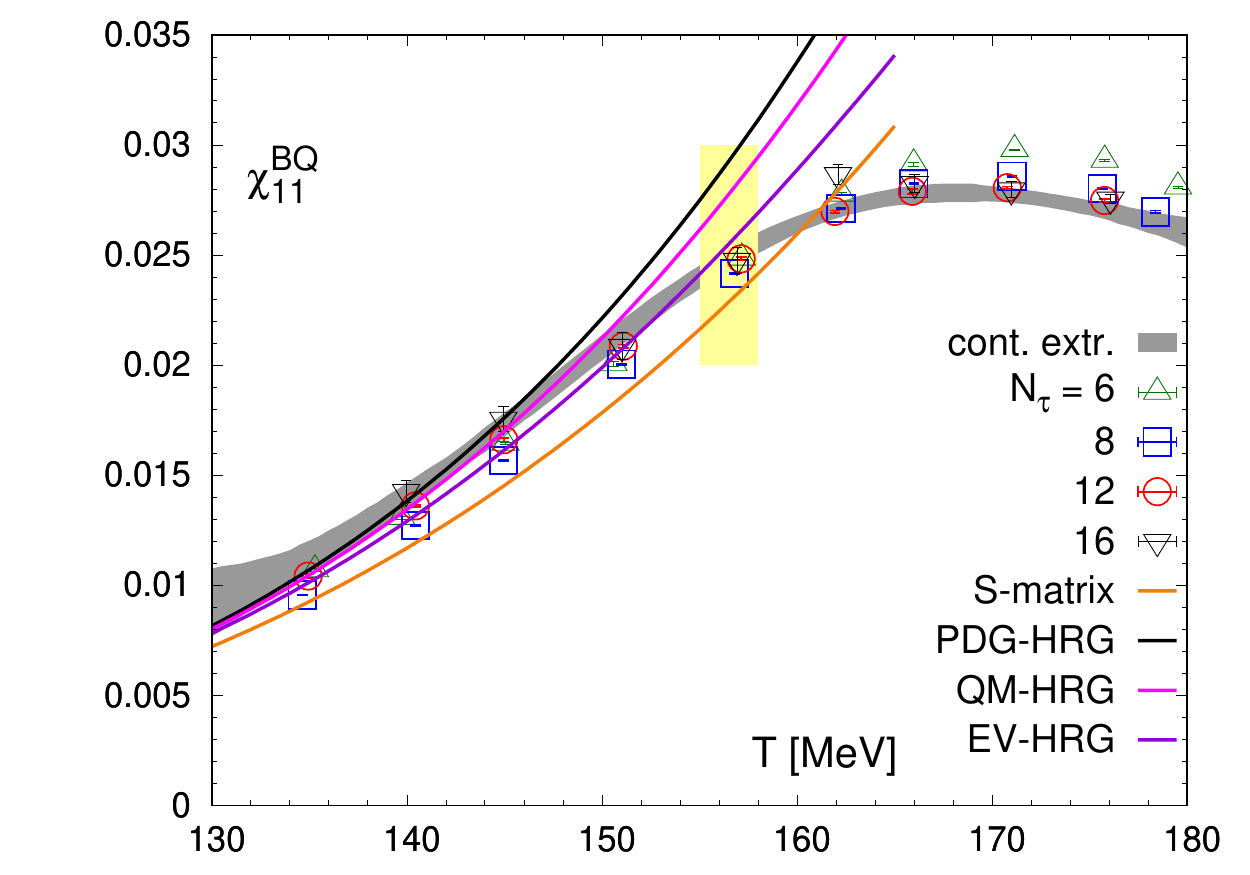}
\includegraphics[width=5cm]{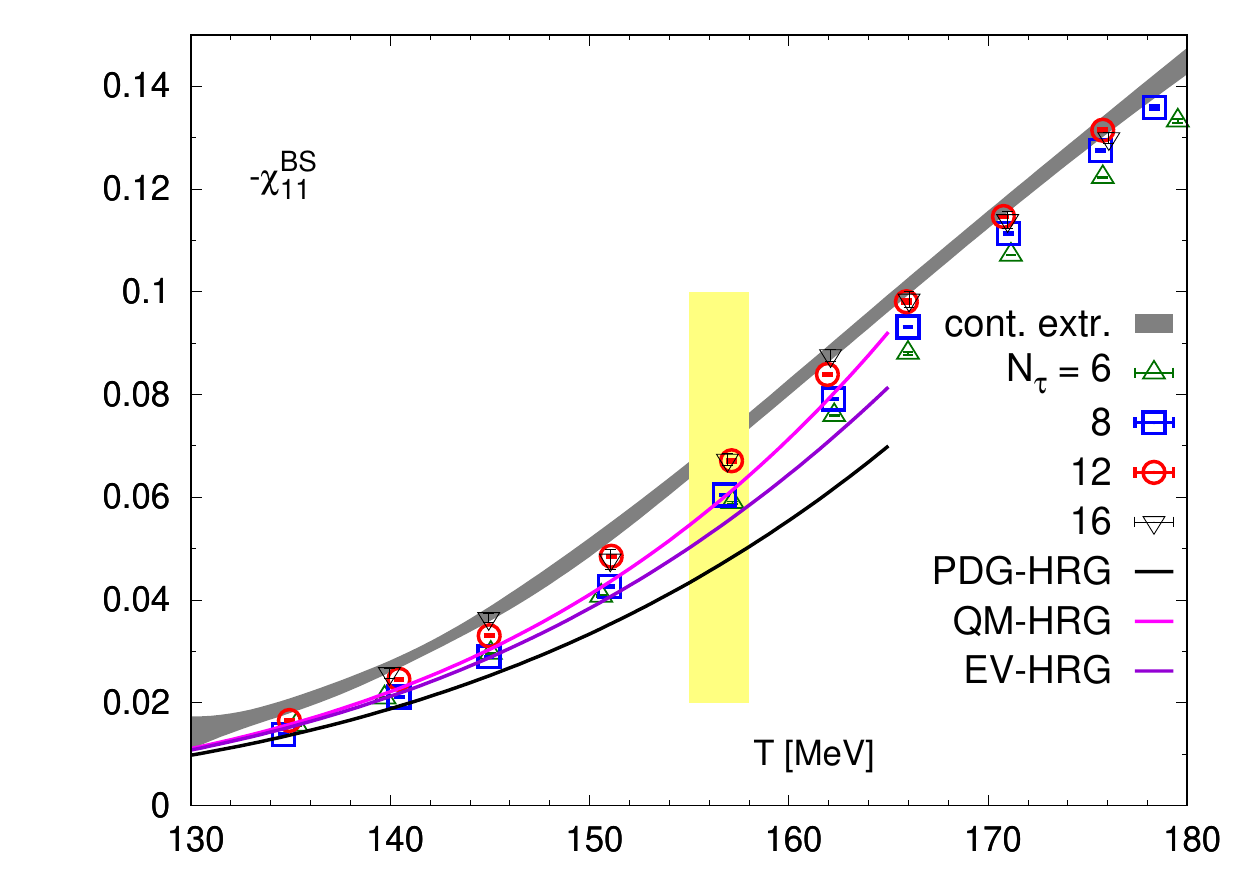}}
\caption{Second order cumulants of net baryon-number fluctuations correlated with net electric charge (BQ) 
and strangeness (BS) fluctuations, respectively.}
\label{fig:BQ_corr}
\end{figure}

At large temperatures the HRG model results are significantly larger than the QCD results. This may, partly, be 
compensated by introducing repulsive interactions in the baryon sector of a HRG. 
When using an excluded volume of size $b\simeq 0.4/T^3$, which corresponds to $rT\simeq 0.3$
or a hadron radius that varies in the range  (0.45-0.34)fm in the temperature range (130-175) MeV
that is of interest in our study\footnote{Note that a constant radius $r\simeq 0.39$~fm has been used
in \cite{Vovchenko:2016rkn,Vovchenko:2017xad}, which is close to our value of $r$ in the pseudo-critical region.},
we find a reduction of $\chi_2^B$ of about 20\% in the transition region. The influence on strangeness fluctuations is much smaller, {\it i.e.} about $8\%$, as these are dominated by mesons. We show results for the QM-HRG with excluded volume effects for baryons (EV-HRG) in Fig.~\ref{fig:2nd_baryon_number}. It is apparent 
that the hadronic interaction considered here is not sufficient to describe the QCD data. They rather tend to
worsen the agreement between HRG and QCD calculations achieved by introducing additional strange baryon
resonances.

\subsection{Charge correlations and constraints on second order cumulants}
In Fig.~\ref{fig:BQ_corr} we show QCD results for correlations among conserved charge fluctuations and compare
with HRG model calculations as discussed above for the $2^{nd}$ order cumulants of conserved charge 
fluctuations. The general picture is the same. Additional strange baryon resonances seem to be needed
to improve agreement between HRG model calculations for $BS$-correlations and corresponding QCD results,
and the inclusion of repulsive interactions among baryons through excluded volume effects seems to 
deteriorate this agreement. Also shown in Fig.~\ref{fig:BQ_corr} is the result of a S-matrix calculation
\cite{Lo:2017lym} that takes into account resonance decays in the $\Delta^{++} \leftrightarrow N^*\pi$ channel
(see also Appendix \ref{AC}).
As can be seen also the S-matrix approach  leads to a reduction of correlations between
net baryon-number and electric charge. The contribution of doubly charged $\Delta^{++}$ resonances thus seems 
to be suppressed.  

\begin{figure}[t]
	\centerline{
		\includegraphics[width=4.4cm]{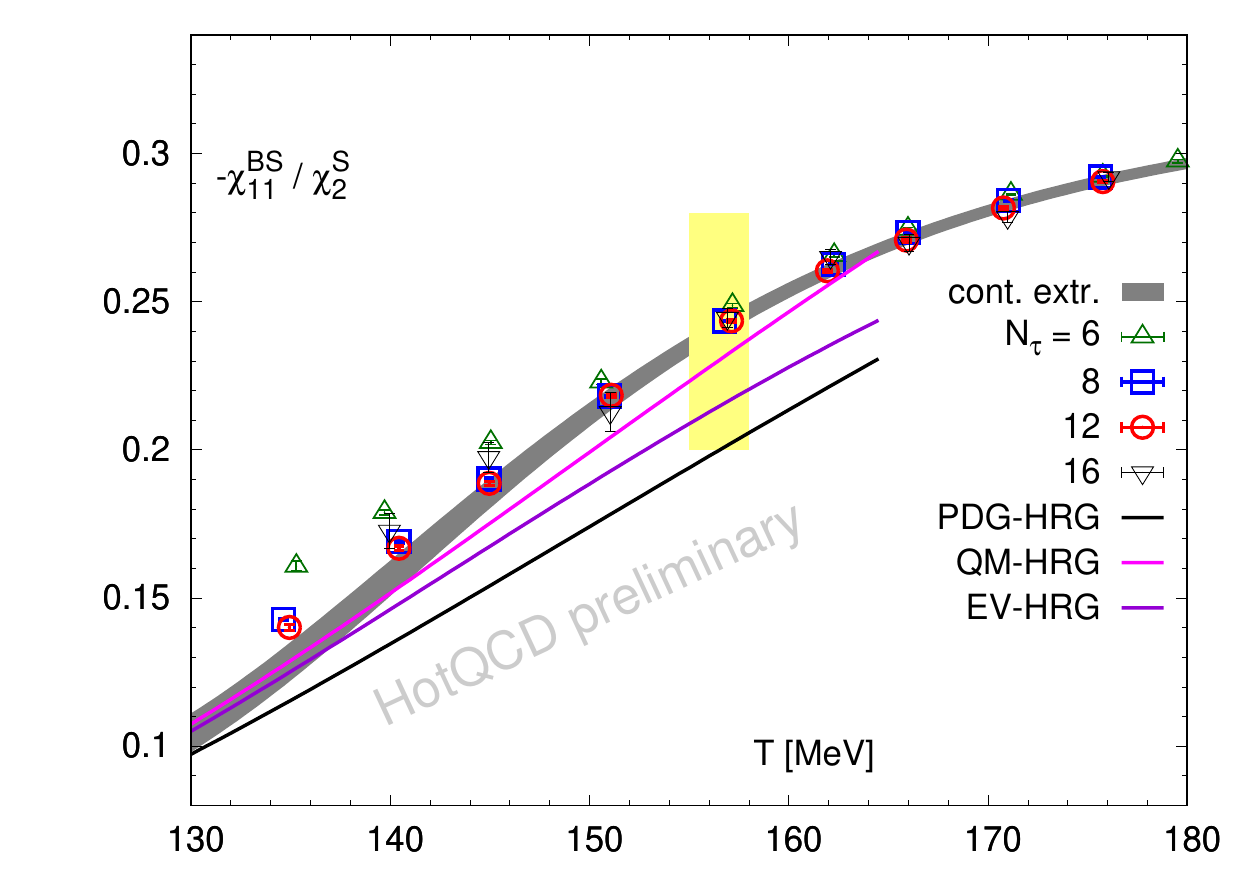}\hspace{-0.3cm}
		\includegraphics[width=4.4cm]{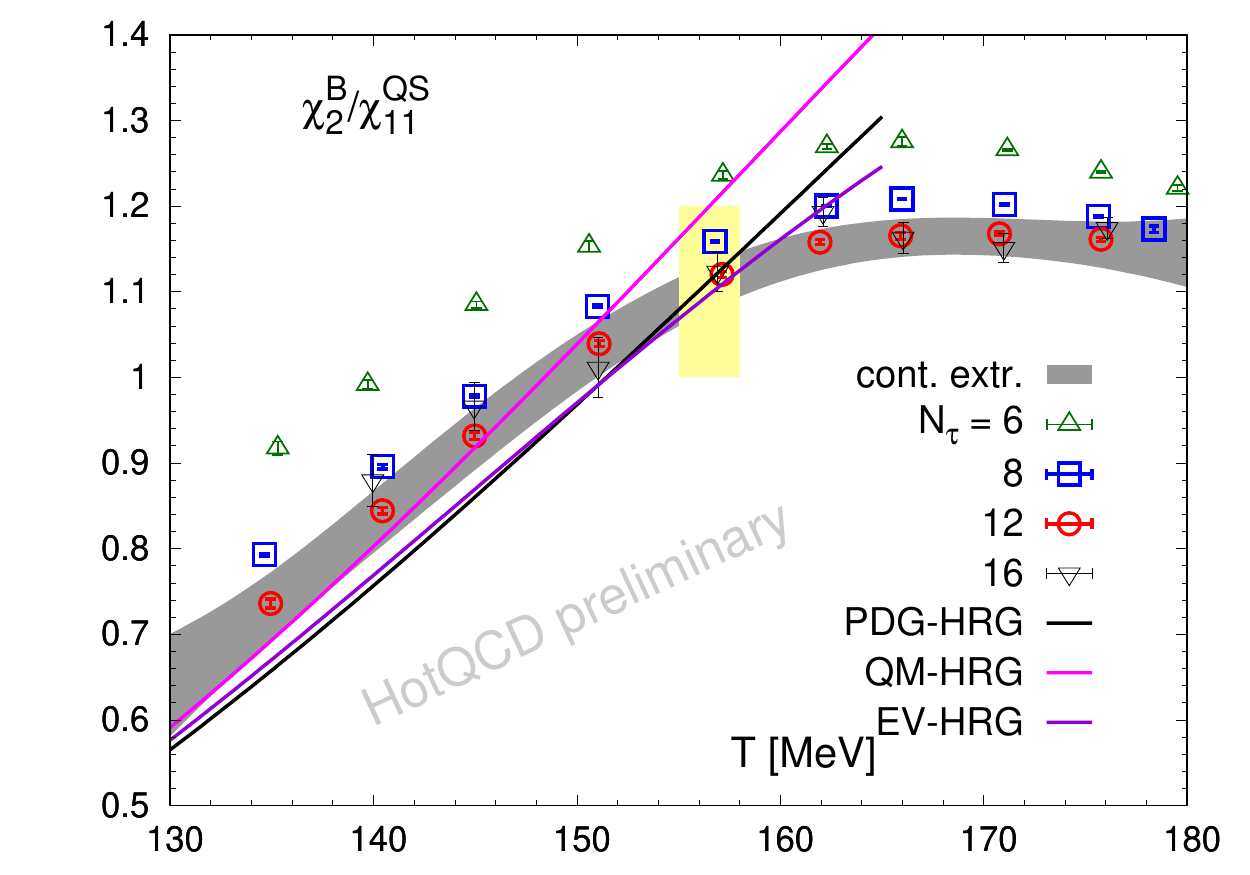}\hspace{-0.3cm}
	\includegraphics[width=4.4cm]{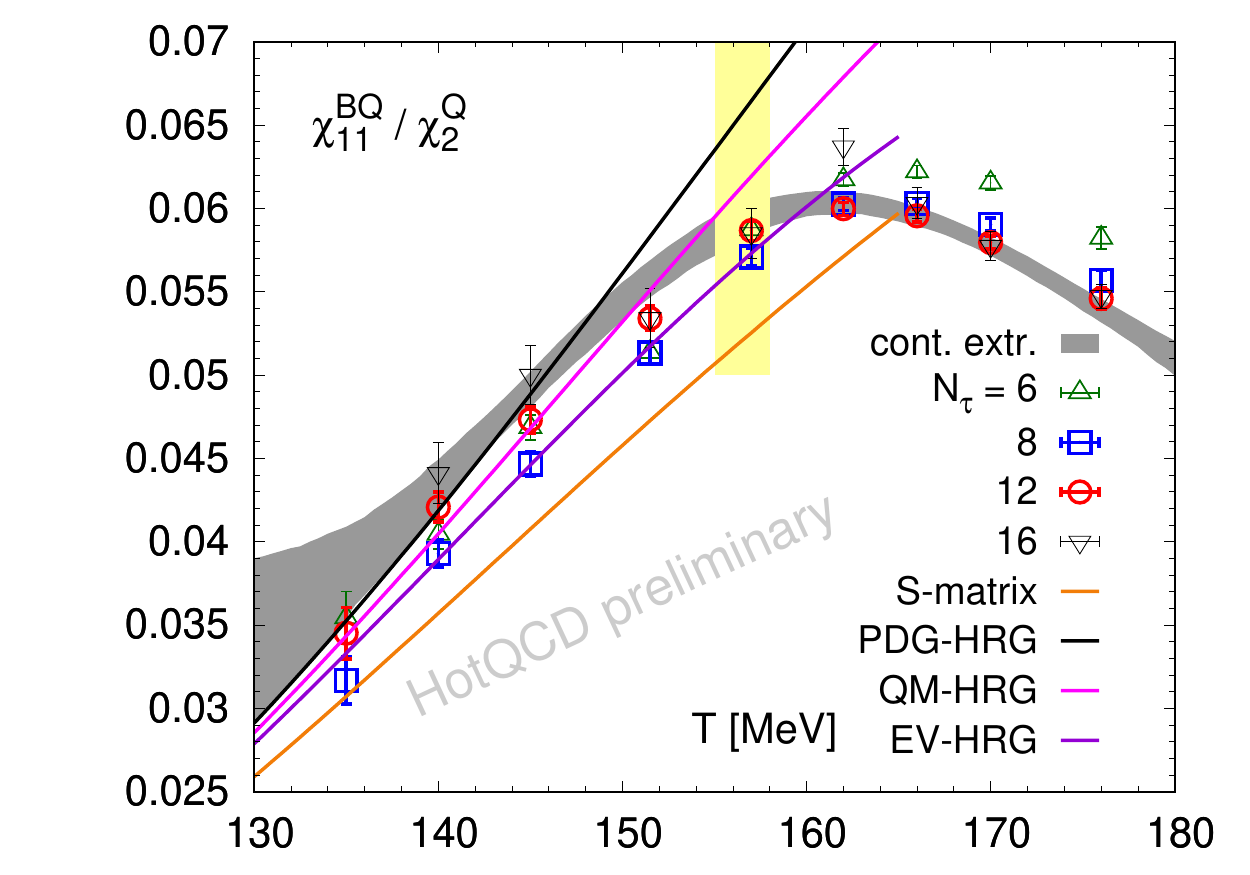}}
	\caption{Continuum extrapolated results for $2^{nd}$ order cumulant ratios.}
	\label{fig:BS_corr_ratio}
\end{figure}

The three conserved charges, $(B,Q,S)$, give rise to 6 second order cumulants of charge fluctuations and 
cross-correlations. In the isospin symmetric limit of degenerate up and down quark masses, which usually
is used in lattice QCD calculations, only 4 of these cumulants are independent as isospin symmetry
imposes the two constraints
\begin{equation}
  \chi_2^S= 2 \chi_{11}^{QS} - \chi_{11}^{BS}\;\; , \; \;   \chi_2^B= 2 \chi_{11}^{BQ} - \chi_{11}^{BS}\; .
\label{constraint}    
\end{equation}
This gives rise to three
independent cumulant ratios, for instance the set of three ratios of second order cumulants 
shown in Fig.~\ref{fig:BS_corr_ratio}. 
In Table~\ref{tab:cumulants2} we give results for continuum extrapolations of these three ratios at the 
pseudo-critical temperature, $T_{pc}= 156.5(1.5)$ MeV, for the chiral transition
in $(2 + 1)$-flavor QCD. Note, for instance, that due to the first constraint in Eq.~\ref{constraint}, the ratio
$\chi^B_2/\chi^S_2$ shown in Fig.~\ref{fig:2nd_baryon_number}~(right) is related to the two ratios $\chi_{11}^{BS}/\chi_2^S$ and $\chi_2^B/\chi_{11}^{QS}$ shown in Fig.~\ref{fig:BS_corr_ratio} and given in Table~\ref{tab:cumulants2} at $T_{pc}$,
\begin{equation}
    \frac{\chi_2^B}{\chi_2^S} = \frac{1}{2} \frac{\chi_2^B}{\chi_{11}^{QS}}\left( 1+ \frac{\chi_{11}^{BS}}{\chi_2^S} \right) \; ,
\end{equation}
and $(\chi_2^B/\chi_2^S)_{T_{pc}}=0.417(18)$.

\begin{table*}[t]
	\centering
	\begin{tabular}{c | c | c } 
		\hline \\[-2ex]
		$\chi_{11}^{BS}/\chi_2^S$ & $\chi_{2}^{B}/\chi_{11}^{QS}$ & $\chi_{11}^{BQ}/\chi_{2}^{Q}$ 
		\\ [1.5ex] 
		\hline \\[-2ex]
		 -0.241(4) & 1.10(4) & 0.059(14) \\ [1ex] 
		\hline
	\end{tabular}
\caption{Continuum extrapolated results for three independent ratios of $2^{nd}$ order cumulants at the pseudo-critical temperature $T_{pc}$.}
\label{tab:cumulants2}
\end{table*}
\begin{figure}[b]
\includegraphics[width=4.4cm]{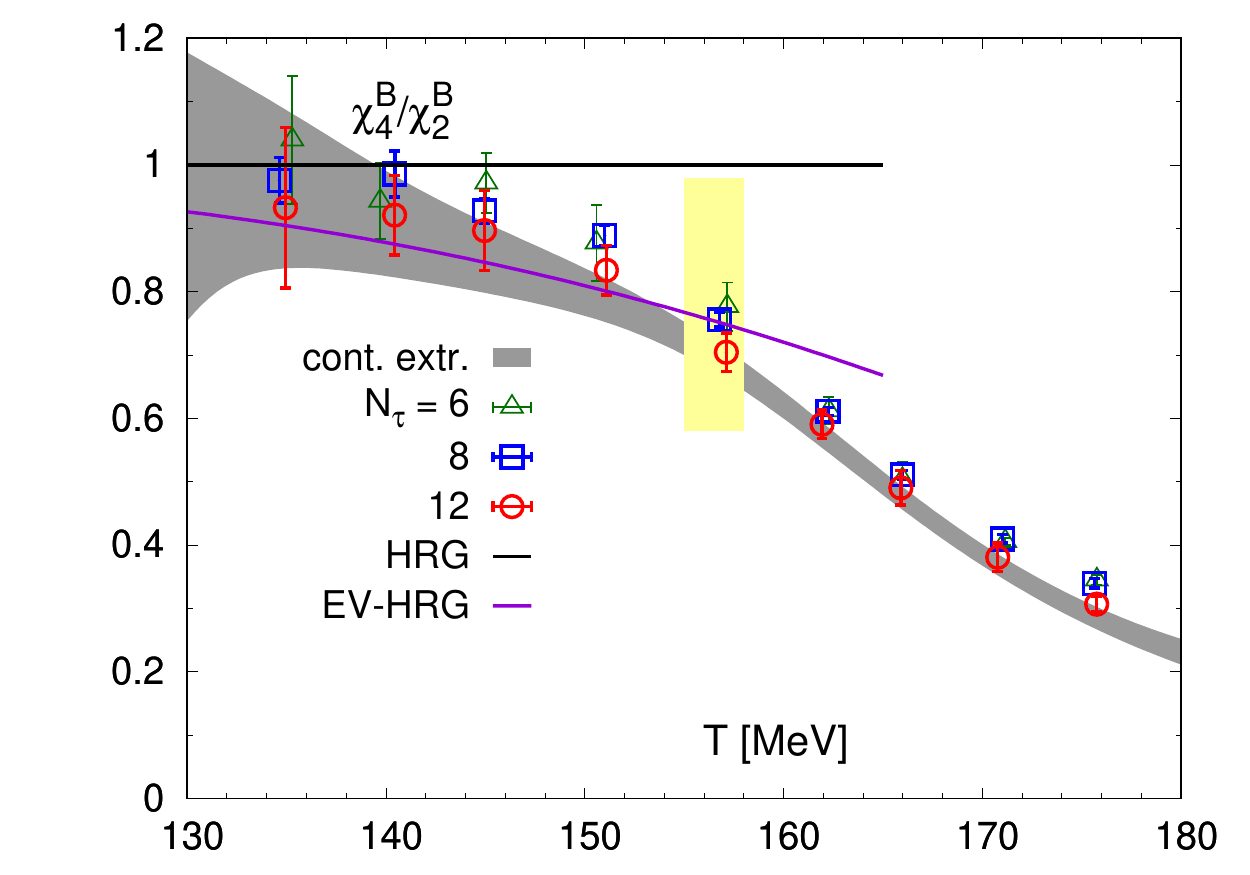}\hspace{-0.4cm}
\includegraphics[width=4.4cm]{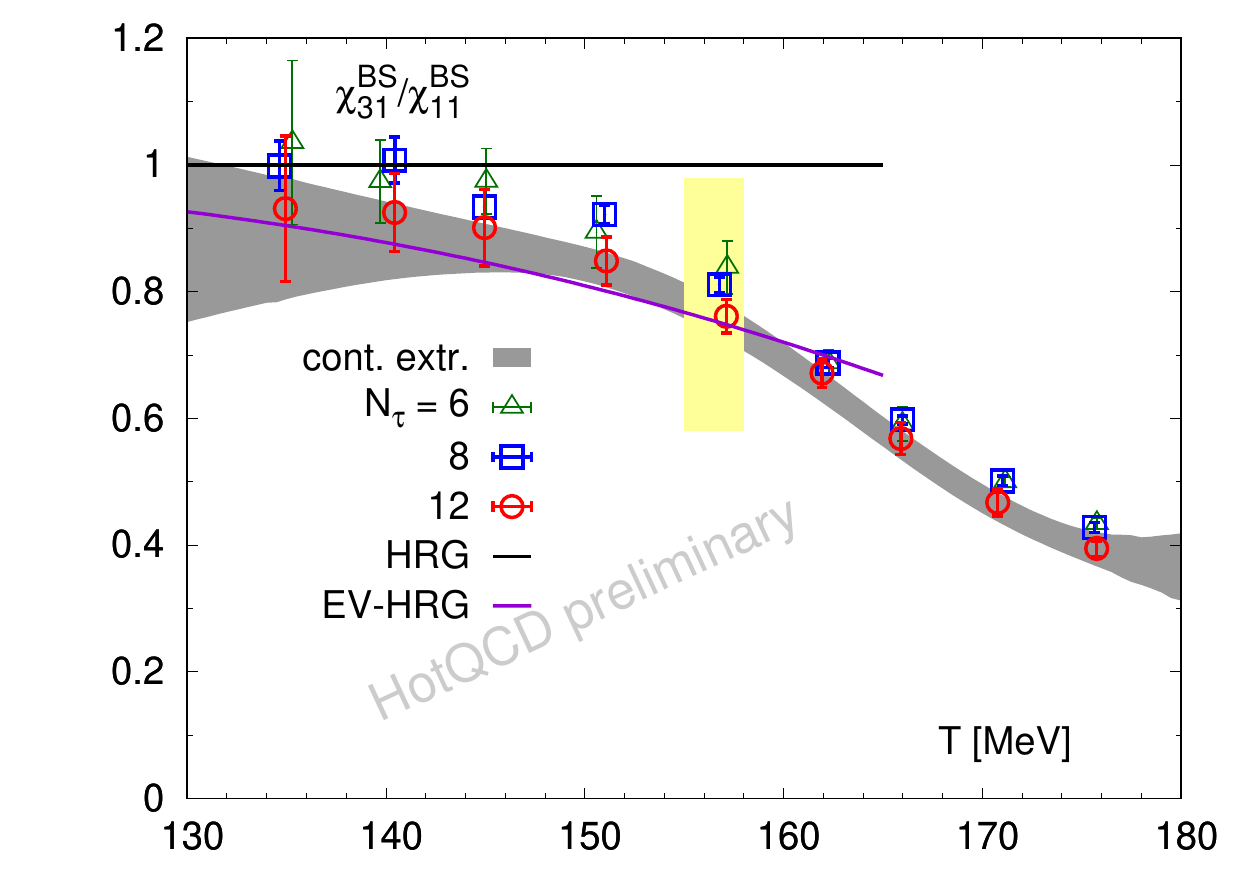}\hspace{-0.4cm}
\includegraphics[width=4.4cm]{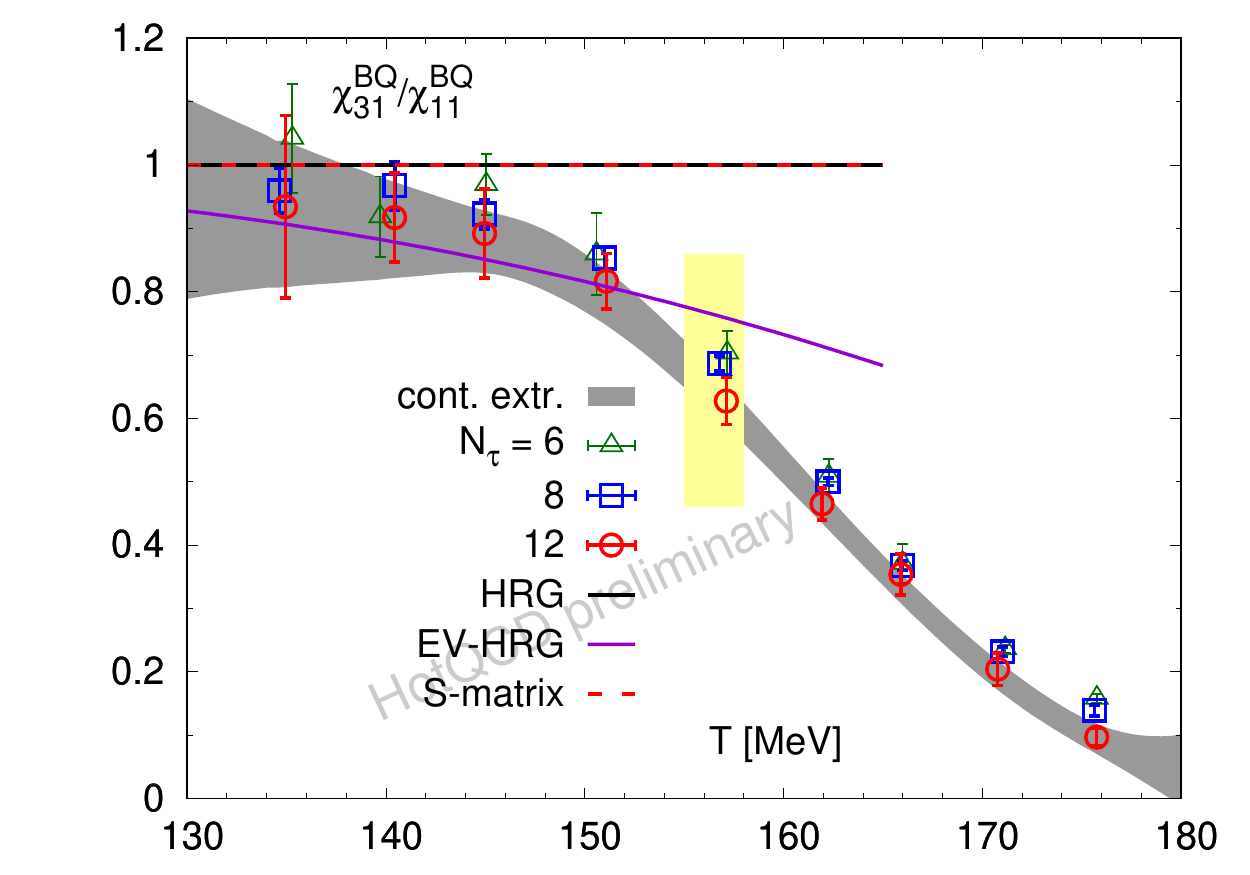}
\caption{Ratios of some fourth and second order cumulants.}
        \label{fig:4th_baryon_number}
\end{figure}

\section{Fourth order cumulants of conserved charge fluctuations and correlations}

In a non-interacting HRG (PDG-HRG or QM-HRG) ratios of cumulants involving net baryon-number fluctuations 
that differ only by an even number of derivatives with respect to the baryon chemical potential are unity, e.g. 
for fourth order cumulants
$\chi_4^B/\chi_2^B=\chi_{31}^{BS}/\chi_{11}^{BS}=\chi_{31}^{BQ}/\chi_{11}^{BQ}=1$. This reflects that all
known hadrons with non-zero baryon number have $|B|=1$. This, of course, does not hold in QCD at high
temperatures where quarks carry non-integer baryon number. As a consequence the above ratios are all found
to be smaller than unity in lattice QCD calculations. They are shown in Fig.~\ref{fig:4th_baryon_number}.

At low temperatures the deviations from unity follow a trend also present in HRG model calculations that 
incorporate excluded volume effects. This also is shown in Fig.~\ref{fig:4th_baryon_number}. Although the agreement
of these model calculations with lattice QCD data seems to be reasonable below the pseudo-critical 
temperature, we note that this is to some extent accidental as the EV-HRG calculations do not provide
an adequate description for neither the numerator nor the denominator of these ratios. E.g., in the case of 
quadratic ($\chi_2^B$) and quartic ($\chi_4^B$) net baryon-number fluctuations the EV-HRG calculations 
both underestimate the QCD results. We also note that in the present implementation of the S-matrix 
approach the ratio $\chi_{31}^{BQ}/\chi_{11}^{BQ}$ remains unity like it is the case in non-interacting
HRG models.

\section{Conclusions}
Qualitative features of $2^{nd}$ and $4^{th}$ order cumulants of conserved charge fluctuations and correlations, calculated in lattice QCD, are reasonably well described by non-interacting HRG models up to the pseudo-critical temperature for the QCD transition. At higher temperatures significant deviations quickly set in, being large
already for temperatures about 10\% higher than $T_{pc}$ and being larger for $4^{th}$ than for 
$2^{nd}$ order cumulants. In order to reach a somewhat satisfactory description of strangeness fluctuations
the addition of strange resonances calculated in quark models (QM-HRG model) is needed, which may be viewed as
taking care of additional interactions, represented for instance in less well-established resonances
listed in the PDG, which are not reflected in HRG models based on down to 3-star resonances only. Including further
interactions through e.g. excluded volume HRG models (EV-HRG) or S-matrix approaches does not seem to lead
to a further quantitative improvement of many of the cumulants considered here.

\section*{Acknowledgements}
This work was supported by the Deutsche Forschungsgemeinschaft (DFG) - project number 315477589 - TRR 211, the European Union H2020-MSCA-ITN-2018-813942 (EuroPLEx), and the U.S. Department of Energy, Office of Science, Office of Nuclear Physics through i) the Contract No. DE-SC0012704 and (ii) within the framework of the Beam Energy Scan Theory (BEST) Topical Collaboration, and (iii) the Office of Nuclear Physics and Office of Advanced Scientific Computing Research within the framework of Scientific Discovery through Advance Computing (SciDAC) award "Computing the Properties of Matter with Leadership Computing Resources''.

\appendix
\section{Non-Interacting HRG Model}
\label{AA}
The pressure of a non-interacting hadron gas can be written as a sum of contributions
for mesons ($M$) and baryons ($B$) and the corresponding anti-particles $(\bar{M},\ \bar{B})$,
\begin{equation}
    P = P_M+P_{\bar{M}}+P_B+P_{\bar{B}} \; , \label{eq:tot_press}
\end{equation}
\begin{eqnarray}
P/T^4 &= &\sum\limits_{H \in B,\bar{B}}\frac{g}{2\pi^2}(m_H/T)^2\sum\limits_{k=1}^{\infty}\frac{(-1)^{k+1}}{k^2}K_2\bigg(\frac{km_H}{T}\bigg)\exp[k\vec{C}_H.\vec{\mu}/T] \nonumber \\ &+ &\sum\limits_{H \in M,\bar{M}}\frac{g}{2\pi^2}(m_H/T)^2\sum\limits_{k=1}^{\infty}\frac{1}{k^2}K_2\bigg(\frac{km_H}{T}\bigg)\exp[k\vec{C}_H.\vec{\mu}/T] \; , \label{eq:ideal_gas_press} 
\end{eqnarray} 
where $\vec{C}_H = (B_H,Q_H,S_H)$ represents the conserved charges, {\it i.e.} baryon number, electric charge and strangeness number of the hadron $H$, and $\mu_B,\mu_Q,\mu_S$ are the baryon, electric charge and strangeness chemical potentials, respectively. $K_2$ is the modified Bessel function of second kind. Generalized susceptibility can be obtained from Eq.(\ref{eq:gen_sus}),
\begin{eqnarray}
\chi_{lmn}^{BQS} &=& \sum\limits_{H \in B,\bar{B}}\frac{g_{_H}}{2\pi^2}(m_H/T)^2~B_H^lQ_H^mS_H^n~K_2\bigg(\frac{m_H}{T}\bigg) \label{eq:non-hrg_sus} \\
&+& \sum\limits_{H \in M,\bar{M}}~\frac{g_{_H}}{2\pi^2}(m_H/T)^2\sum\limits_{k=1}^{\infty}(kQ_H)^m(kS_H)^n\frac{1}{k^2}K_2\bigg(\frac{km_H}{T}\bigg) \nonumber
\end{eqnarray}
In Eq.(\ref{eq:non-hrg_sus}) the first term corresponds to the baryon sector, where we only used the Boltzmann
approximation to the Fermi sum given in Eq.~\ref{eq:ideal_gas_press},
and the second term corresponds to the meson sector. Note that the second term will drop out from the Eq.(\ref{eq:non-hrg_sus}) for any baryonic observables  as baryon number($B$) is 0 for mesons.

\section{Excluded Volume HRG Model}
\label{AB}
In excluded volume we only consider the interaction between baryons, $BB$, and anti-baryons, $\bar{B}\bar{B}$. 
The meson-meson, $MM$, and meson-baryon, $MB(\bar{B})$, as well as baryon-antibaryon, $B\bar{B}$, interactions
are neglected. Hence, the excluded volume will only modify $P_{B}$ and $P_{\bar{B}}$ independently and the total pressure Eq.(\ref{eq:tot_press}), can be replaced by, 
\begin{eqnarray}
P=P_M+P_{\bar{M}}+P_B^{int}+P_{\bar{B}}^{int} \; , \label{eq:tot_press_excl}
\end{eqnarray}
where the interacting baryon or anti-baryon pressure can be written as,
\begin{eqnarray}
\hat{P}^{int}_{B/\bar{B}} &=& \sum_{H \in B/\bar{B}} \hat{P}^{id}_H(T,\vec{\mu})
~\exp[- b^{\prime}\hat{P}^{int}_{B/\bar{B}}], \\
\label{eq:excl_vol}
\end{eqnarray}
with $ b^{\prime}=bT^3$ and $\hat{P}=P/T^4$. This equation may be solved iteratively,
\begin{eqnarray}
\hat{P}^{int}_{B/\bar{B}} &=&\sum_{H \in B/\bar{B}}\hat{P}_H^{id}(T,\vec{\mu}) - b^{\prime} \bigg[\sum_{H,H^{\prime}\in B/\bar{B}}\hat{P}^{id}_{nm}(T,\vec{\mu})\bigg]^2 \\
&&+ (3{b^{\prime}}^2/2)\bigg[\sum_{H,H^{\prime},H^{\prime\prime}\in B/\bar{B}}\hat{P}^{id}_{lmn}(T,\vec{\mu})\bigg]^3 
                                   +.....\nonumber
\label{eq:excl_volexp}
\end{eqnarray}
For the baryon species $H$, $\hat{P}^{id}_H(T,\vec{\mu})$ can be written from Eq.(\ref{eq:ideal_gas_press})  using Boltzmann approximation as,
\begin{eqnarray}
\hat{P}^{id}_H &= & \frac{g_{_H}}{2\pi^2}(m_H/T)^2 K_2\bigg(\frac{m_H}{T}\bigg)\exp[\vec{C}_H.\vec{\mu}/T].
\end{eqnarray}
The term linear in $b^{\prime}$ appearing in Eq.(\ref{eq:excl_volexp}) acts as a repulsive term. It decreases the pressure 
and is related to the second virial coefficient. The term quadratic in $b$  acts as an attractive term. 
However, since $\hat{P}^{id}_H\sim \exp(-m_H/T)$, only the term linear in $b$ will survive for $m_H\gg T$ at high temperature, {\it i.e.} excluded volume effects are predominantly repulsive. Moreover, since for low temperatures $\hat{P}^{id}_H\to0$, 
$\hat{P}^{int}_H$ will also approach $\hat{P}^{id}_H$. 

Generalized susceptibilities can be obtained by taking derivatives of Eq.(\ref{eq:excl_vol}) with respect to chemical potentials ($\mu_B,\mu_Q~\rm{and}~\mu_S$).
We also note that the resulting equations in the mean field approach \cite{Huovinen:2017ogf} are quite similar to 
those obtained in the excluded volume approach. The difference is that in mean field approach the repulsive 
interactions have between used only for the baryon octet and decuplet, while in the excluded volume approach one generally considers interaction between all baryons.

\section{S-matrix formalism}
\label{AC}
In the S-matrix formalism as used here by us, we only considered the decay and production of 
$N^* \rm{and}~\Delta^{++}$ to pion and nucleon in a hot hadron gas. The pressure can then be separated into two 
parts; one part arises from the interaction of pion and nucleon, the other part describes the contribution from
all other particles,
\begin{eqnarray}
\hat{P}&=&\hat{P}^{id}+\sum_{I_z=3/2,1/2}\hat{P}^{int}\; , \\
\hat{P}^{int} &=& \frac{g}{2\pi^2}~\int_{m_{th}}^{\infty}~d\epsilon~K_2(\epsilon/T)~(\epsilon/T)^2~\frac{d\delta_{IJ}}{\pi d\epsilon} \; ,
\end{eqnarray}
where $\hat{P}^{id}$ is same as in Eq.(\ref{eq:ideal_gas_press}), but without the contribution from $\Delta^{++}$ and $N^*$ resonances. Their contribution to the pressure is included in the $\hat{P}^{int}$.
Here we follow the notation and steps of \cite{Venugopalan:1992hy,Lo:2017lym} for calculating the cumulants
of net baryon-number and electric charge correlations, $\chi_{nm}^{BQ}$, in the S-matrix approach.

\end{document}